\documentclass[12pt]{article}
\usepackage{cite}

\newcommand{\Tr}{{\rm Tr}\,}
\newcommand{\ntwo}{\mbox{${\cal N}\!\!=\!2\;$}}
\newcommand{\none}{\mbox{${\cal N}\!\!=\!1\;$}}

\begin{document}
\begin{titlepage}

\begin{flushright}
TPI-MINN-01/25 \\
UMN-TH-2007/01\\
hep-th/0106176
\end{flushright}

\vfil

\begin{center}
{\large\bf
Monopole, gluino and charge condensates in gauge \\[2mm] 
theories with 
broken ${\cal N}$=2  supersymmetry$^{\, 1}$}

\vspace{0.4in}

Arkady Vainshtein$^{\, 2}$

\vspace{0.2in}

{\it Theoretical Physics Institute, University of Minnesota,\\
Minneapolis, MN 55455}

\end{center}

\vfil

\begin{center}
{\bf Abstract}
\end{center}

\vspace{2mm}

We consider chiral condensates in  SU(2) gauge theory 
with broken \ntwo su\-persymmetry and one  fundamental flavor in the matter
sector.  Matter and gaugino condensates 
are determined by integrating out the adjoint field. The only
nonperturbative  input  is the Affleck-Dine-Seiberg one-instanton  superpotential.
 The results are
consistent with those obtained by the `integrating in' procedure.
 We then  calculate monopole, dyon, and charge
condensates using the Seiberg-Witten approach. The key observation is that the
monopole and charge condensates  vanish at  the Argyres-Douglas point where
the monopole and charge vacua collide. We interpret this phenomenon
as  a deconfinement of electric and magnetic charges at  the
Argyres-Douglas  point. 

\footnotetext[1]{\footnotesize
Talk presented at the Conference ``Quantization, Gauge Theory, and Strings''
dedicated to the memory of Professor Efim Fradkin, Moscow, Russia, 5-10 June
2000, to appear in the Proceedings.}
\footnotetext[2]{\footnotesize
The work is  supported in part
by DOE under the grant DE-FG02-94ER40823.}
\setcounter{footnote}{2}
\end{titlepage}

\section{Introduction}

This talk is based on
the work \cite{GVY} in which I have collaborated with Alexander Gorsky and
Alexei Yung.

 The derivation of exact results in \none supersymmetric
gauge theories from low energy effective superpotentials and holomorphy was
pioneered in \cite{ads,SV}. Then a new wave of development was initiated by
Seiberg, see~\cite{IS} for review.  Additional  input was provided by the
Seiberg-Witten solution 
 of \ntwo supersymmetric gauge theories with and without
matter~\cite{sw}.

In the \ntwo theory  the chiral \none su\-per\-field $\Phi$
in  the adjoint representation appears as a partner  to the gauge fields in 
the \ntwo
super\-multiplet.  The key feature of the \ntwo theory  is the existence of the
Coulomb branch where the  vacuum expectation value of the lowest
component of $\Phi$  serves as a modulus \cite{sw}.

The simplest way to break  \ntwo supersymmetry (SUSY)
 down to \none am\-o\-unts to
giving  a nonvanishing   mass $\mu$ to the  su\-per\-field $\Phi$.
 At small
values of  $\mu$ the theory is close to its \ntwo counterpart while at
large  $\mu$ the adjoint matter decouples and the pure \none theory
emerges.  The emerging  theory at large $\mu$ is close to 
supersymmetric  QCD (SQCD) but  does not coincide with 
it. A trace of the massive adjoint remains in the effective theory in the
 form of nonrenormalizable quartic terms~\cite{kutasov} in the superpotential
which are  suppressed by $1/\mu$.
Although in the \none theory the degeneracy on the Coulomb branch is lifted by
the superpotential, 
 memory of the structure of the Riemann surfaces remains.
Namely, the vanishing of the 
discriminant of the  Riemann surface defines the set of vacua in
the corresponding \none theory~\cite{sw,kutasov,is,giveon,kitao,kt}. 

We consider an \none theory with both adjoint and fundamental
matter and limit ourselves to the most
 tractable case of SU(2) gauge group with
one fundamental flavor and one multiplet in the adjoint representation.
Our strategy is as follows.  First, in Sec.~\ref{sec:sup}, we integrate 
out the adjoint
matter to get 
 SQCD-like effective superpotential for the fundamental matter. The only
nonperturbative input in this effective superpotential is given by the   
Affleck-Dine-Seiberg (ADS) superpotential generated by one instanton 
\cite{ads}.  Difference 
with  pure SQCD is due to the mentioned above nonrenormalizable term
generated by the level  heavy adjoint exchange. Similarly to SQCD, the 
effective 
superpotential together with the Konishi relations unambiguously fixes 
condensates
of fundamental and  adjoint matter as well as the gaugino condensates in 
all three
vacua of the theory.

Our results for matter and gaugino condensates are consistent with those 
obtained by the `integrating in' method~\cite{ils,Intrilligator,giveon} 
and can be
viewed  as an independent confirmation of this method. 
What is specific to our approach is that
we start from the  weak coupling regime where the notion of an effective
Lagrangian is  well defined, and  then use holomorphy to extend results 
for chiral condensates into strong coupling. 

In Sec.~\ref{sec:mcd} we  determine monopole, dyon, and charge 
condensates following  the Seiberg-Witten approach, i.e. considering 
effective superpotentials near singularities on the Coulomb branch of the 
 \ntwo theory. Again, holomorphy allows us to extend our results to the
domain of the ``hard'' \ntwo breaking. This extension includes not only
the mass term of adjoint but also  breaking of \ntwo in Yukawa couplings.

Our particular interest  is the study of chiral condensates  in the 
Argyres-Douglas
(AD) points. These points were originally introduced in the moduli/parameter
space of \ntwo theories as 
points  where two singularities on the Coulomb branch
coalesce~\cite{ad,apsw,hori}. It is believed that the theory in
 the AD point flows in the infrared to a nontrivial
superconformal theory. The notion of the AD point continues to
make sense even when the  \ntwo theory  is broken to \none; in
the \none theory it is the point in parameter space where 
 two vacua collide.

In particular, we consider the AD point where the monopole and charge
vacua collide at a certain
 value of the mass of the fundamental flavor. Our key result is
that  both monopole and charge condensates  vanish at the
 AD  point\,\footnote{Vanishing of condensates for coalescing vacua was
   mentioned by Douglas and Shen\-ker~\cite{DS} in the context of SU($N$)
  theories without fundamental matter for $N\ge 3$. Note, that it
  was even before the notion of the AD point was introduced in~\cite{ad}. }.  
We interpret this  as deconfinement of both
electric and magnetic charges at the  AD
point. It provides evidence that the theory
at the AD point remains superconformal even after strong breaking of
\ntwo to ${\cal N}\!\!=\!1$. Argyres and Douglas conjectured this in their
consideration of SU(3) theory~\cite{ad}. 

\section{Matter and gaugino  condensates and effective superpotential}
\label{sec:sup}

We consider a \none theory with SU(2) gauge group where the matter
sector consists of the adjoint field $\Phi^\alpha_\beta=\Phi^a
(\tau^a/2)^\alpha_\beta$ ($\alpha,\beta=1,2;~a=1,2,3$), and two fundamental 
fields
$Q^\alpha_f$ $(f=1,2)$ describing one flavor.
The  general renormalizable superpotential for this theory has the form,
\begin{equation}
{\cal{W}}= \mu \,{\rm Tr} \,\Phi^2 + \frac{m}{2} \,Q_{ f}^\alpha Q^{f}_\alpha  
+\frac{1}{\sqrt{2}}\,h^{fg} \,Q_{\alpha f}\Phi^\alpha_\beta Q^\beta_g\;.
\label{superp}
\end{equation}
Here the parameters $\mu$ and $m$ are related to the masses of the adjoint and
fundamental fields, 
$m_\Phi=\mu/Z_\Phi$, $m_Q=m/Z_Q$, by the corresponding $Z$ factors in 
the kinetic
terms. Having  in mind normalization appropriate for
 the \ntwo case we choose for bare parameters
$Z_\Phi^0=1/g_0^2$, $Z_Q^0=1$. The matrix of Yukawa couplings
 $h^{fg}$ is  symmetric, and  summation over color indices $\alpha,\beta=1,2$
is explicit. Unbroken \ntwo SUSY appears when $\mu=0$ and $\det h=-1$.

To obtain an effective theory similar to SQCD
 we integrate out the adjoint field
$\Phi$ implying  that $m_\Phi\gg m_Q$. In the classical approximation this 
integration 
reduces to  the substitution
\begin{equation}
\Phi^\alpha_\beta= -\frac{1}{2\sqrt{2}\,\mu}
\,h^{fg}\left( Q_{\beta f}Q^\alpha_g
-\frac 1 2\, \delta^\alpha_\beta \, Q_{\gamma f}Q^\gamma_g\right)\,,
\label{substitute}
\end{equation}
which follows from $\partial {\cal{W}}/\partial \Phi =0$.  What is the effect
 of
quantum corrections on the effective superpotential? It is well known from the
study of SQCD that perturbative loops do not contribute and nonperturbative 
effects
are exhausted  by the  Affleck-Dine-Seiberg (ADS)
superpotential generated by one instanton~\cite{ads}. 
The
effective superpotential then is
\begin{equation}
{\cal{W}}_{\rm eff}= m\, V - \frac{(-\det h)}{4\mu}\, V^2 +\frac{\mu^2
\Lambda_1^3}{4\,V}
\label{sup1}
\end{equation}
where the gauge and subflavor invariant chiral field $V$ is defined as
\begin{equation}
V=\frac 1 2 \,Q_{ f}^\alpha Q^{f}_\alpha \;.
\end{equation}
The first two terms in  Eq.~(\ref{sup1}) appear
 on the tree level after substitution~(\ref{substitute}) into 
Eq.~(\ref{superp})  while the third nonperturbative one is the  ADS
superpotential.  The scale parameter
$\Lambda_1$ is given in terms of the mass of 
Pauli-Villars regulator $M_{\rm PV}$ and
the bare coupling $g_0$ (plus the vacuum angle $\theta_0$) as
\begin{equation}
\Lambda_1^3=4\, M_{\rm PV}^3\exp\left(-\frac{8\pi^2}{g_0^2}+i\theta_0\right)\;.
\end{equation}
The coefficient $\mu^2 \Lambda_1^3/4$ in the ADS superpotential is 
equivalent to $\Lambda^5_{\rm SQCD}$ in SQCD.  The factor $\mu^2$ in the
coefficient reflects four fermionic  zero modes of 
the adjoint field. 

The only term in the superpotential~(\ref{sup1}) which differentiates it from
the SQCD case  is  the second term  which is due to tree level
exchange by the adjoint field. At 
$h=0$ it vanishes and we are back to the known SQCD case with two vacua and 
a Higgs phase for small $m$. 

When $\det h$ is nonvanishing we have three vacua, marked by the vevs of 
the lowest component of $V$,
\begin{equation}
v=\langle \,V \,\rangle\,.
\end{equation}
These vevs are roots of 
 the algebraic equation ${\rm d}{\cal W}_{\rm eff}/{\rm d}v=0$ which 
has the form
\begin{equation}
m-\frac{(-\det h)}{2}  \, \frac{v}{\mu}- \frac{\Lambda_1^3}{4} \left(
\frac{\mu}{v}\right)^2=0\;.
\label{vaceq}
\end{equation}
This equation shows, in particular, that although the second term in the
superpotential (\ref{sup1}) seems to be suppressed at large $\mu$ 
it turns out to be   of the same
order as the ADS term. From Eq.~(\ref{vaceq}) it is also clear that the 
dependence
on
$\mu$ is  given by the scaling $v\propto\mu$. 

To see the dependence on the other
parameters let us
substitute $v$ by the dimensionless variable
$\kappa$ defined by the relation 
\begin{equation}
v=\mu\,\sqrt{\frac{\Lambda_1^3}{4m}}\,\kappa\;.
\label{vk}
\end{equation}
Then Eq.~(\ref{vaceq}), when rewritten in terms of $\kappa$,
\begin{equation}
1-\sigma\, \kappa -\frac{1}{\kappa^2}=0
\label{kappa}
\end{equation}
is governed by the dimensionless parameter $\sigma$,
\begin{equation}
\sigma= \frac{(-\det h)}{4} \,\left(\frac{\Lambda_1}{m}\right)^{3/2}.
\label{sigma}
\end{equation}
We see that the two parameters $m$ and $\det h$ enter only as $m\,(-\det
h)^{-2/3}$. The dependence of $v$ on $\mu$ is linear as we discussed above. 

The particular dependence of condensate $v$ on the parameters
 $\mu$, $m$ and $\det h$  follows from
the  $R$ symmetries of the theory. Following
Seiberg~\cite{seibR} one can consider $\mu$,  $m$ and $\det h$ 
 as background  fields and identify  two nonanomalous
$R$ symmetries which prove the dependence discussed above.
 The charges
of the fields and parameters of the theory under these two U(1) symmetries
are shown in Table 1.
\begin{table}[h]
\begin{center}
\begin{tabular}{|c|c|c|c|c|c|c|c|}
\hline
~ & ~  & ~  &~ & ~ & ~ & ~ & ~ \\[-0.1cm]
{\rm Fields/parameters} &  $\Phi$ & $ Q$ &  $W$ &$\theta$ &$ m$ & $\mu$ & $h$ 
 \\[0.2cm]
\hline
\vspace*{-0.2cm}
~ & ~  & ~  &~ & ~ & ~ & ~ & ~\\
U$_{J}(1)~{\rm charges}$ & 0 & $1$ & $1$   & $1$ & 0   & $2$   
 &0\\[0.2cm]
\hline
\vspace*{-0.2cm}
~ & ~  & ~  &~ & ~ & ~ & ~ & ~\\
U$_{R}(1)~{\rm charges} $ & 1 & $-1$ & 1   & 1 & 4   & 0       &3\\[0.2cm]
\hline
\end{tabular}
\caption{Nonanomalous U(1) symmetries}
\label{tabU}
\end{center}

\vspace*{-6mm}

\end{table}
The first of these symmetries U$_{J}(1)$ is a subgroup
of the global SU$_{R}(2)$ group related to the ${\cal N}=2$
superalgebra~\cite{sw}. The second nonanomalous symmetry U$_{R}(1)$ is similar
to the $R$ symmetry of Ref.~\cite{ads} extended to include the adjoint field. 
As a consequence, for a given chiral field $X$
\begin{equation}
  \label{mdep}
  \langle X \rangle = \mu^{Q_J/2} m^{Q_R/4}\Lambda_1^{d_X
    -(Q_J/2)-(Q_R/4)} f_X(\sigma)\;, 
\end{equation}
where ${Q_J}$,  ${Q_R}$ are the U$_J (1)$, U$_R (1)$ charges of the
field $X$, $d_X$ is its dimension, and $f_X$ is an arbitrary function 
of the dimensionless parameter $\sigma$ defined by Eq.~(\ref{sigma}).

The important benefit of the consideration above is that in a theory 
with \ntwo SUSY strongly broken by large $\mu$
and $\det h\! \neq\! -1$ we  can still relate chiral condensates with those
in softly broken 
\ntwo where $\det h=\!-1$ and  $\mu$ is small.

Here is an example.
When $\sigma\!\to\! 0$ two roots of Eq.\ (\ref{kappa}) are $\kappa_{1,2}=\pm1$
and  the third one goes to infinity as $\kappa_{3}\!=\! 1/\sigma$.
For two finite roots one can suggest dual interpretations.
 Firstly, taking $h\!=\!0$,
one can relate them to two vacua of SQCD in the Higgs phase. Second, for $\det
h\!=\!-1$ (which is its \ntwo value) one can make  $\sigma$ small by taking the
limit of large $m$. But this limit should bring us  to the monopole and dyon
vacua of softly broken \ntwo SYM. The naming of vacua refers to the particle 
whose
mass vanishes in the corresponding vacuum.

To verify this interesting mapping we need to determine the  vev  
\begin{equation}
u=\langle U\rangle=\langle{\rm Tr} \,\Phi^2\rangle\;,
\end{equation}
which can be accomplished using the set of Konishi anomalies. 
Generic equation for an arbitrary matter field $Q$
 looks as follows:
\begin{equation}
\frac{1}{4}\,\bar{D}^{2}J_Q= Q\,\frac{\partial \,{\cal{W}}}{\partial
Q}+T(R)\,\frac{{\rm Tr}\,W^2}{8\pi^2}\;,
\end{equation}
where $T(R)$ is the Casimir in the matter representation. The left
hand side is a total derivative in superspace so its average over 
any supersymmetric
vacuum vanishes. In our case this
 results in two relations for the  condensates,
\begin{eqnarray}
\left\langle \frac{m}{2}\,Q_{ f}^\alpha Q^{f}_\alpha +
\frac{1}{\sqrt{2}}\,h^{fg}
\,Q_{\alpha f}\Phi^\alpha_\beta Q^\beta_g+
\frac12 \,\frac{{\rm Tr}\,W^2}{8\pi^2}
\right\rangle=0 \nonumber\\[2mm]
\left\langle 2\,\mu {\rm Tr} \,\Phi^2 + \frac{1}{\sqrt{2}}\,h^{fg} \,Q_{\alpha
f}\Phi^\alpha_\beta Q^\beta_g +2\,\frac{{\rm Tr}\,W^2}{8\pi^2} 
\right\rangle=0
\label{konishirel}
\end{eqnarray}
From the first relation, after the 
substitution in (\ref{substitute}) and comparing 
with
Eq.~(\ref{vaceq}), we find an expression for gluino condensate~\cite{NSVZ}
\begin{equation}
s=\frac{\langle {\rm Tr}\,\lambda^2\rangle}{16\pi^2}=-\frac{\langle {\rm
Tr}\,W^2\rangle}{16\pi^2}=\frac{\mu^2 \Lambda_1^3}{4\,v}\;.
\end{equation}
This is consistent with the general expression $[T_G -\sum T(R)]\langle {\rm
Tr}\lambda^2\rangle/16\pi^2$ for the nonperturbative ADS piece of the
superpotential (\ref{sup1}), see~\cite{ds}. Combining the two relations in
(\ref{konishirel}) we can express  the condensate  $u$ in terms of $v$, 
\begin{equation}
u=\frac{1}{2\mu}\left(m\,v +3\,s\right)=
\frac{1}{2\mu}\left(m\,v +\frac 3 4\,\frac{\mu^2 \Lambda_1^3}{v}
\right)=
\frac{\sqrt{m\Lambda_1^3}}{4}\left(\kappa+\frac{3}{\kappa}\right)\;.
\label{ukappa}
\end{equation}
Now we see that in the limit of large $m$ two vacua $\kappa=\pm 1$ are in 
perfect
correspondence with $u=\pm \,\Lambda_0^2$ for the monopole and dyon vacua of
\ntwo SYM. Indeed,
$\Lambda_0^4=m\Lambda_1^3$ is the  correct relation between the 
scale parameters of the theories.

The opposite limit of massless fundamentals $m \to 0$ corresponds to $\sigma \to 
\infty$.  In this limit the 
three vacua are related by a $Z_3$  symmetry~\cite{sw},
\begin{equation}
v=\frac{\mu\,\Lambda_1}{(2\det h)^{1/3}}\,e^{2\pi i k/3}\,,\qquad
u= \frac 3 8 \, \Lambda_1^2 \left(2\det h\right)^{1/3}\,e^{-2\pi i k/3}\,.
\label{smc}
\end{equation}
where $k=0,\pm1$ marks different vacua.
Note that the  massless limit exists due to the nonvanishing Yukawa coupling.
When $h\to 0$ we are back to the runaway vacua of massless SQCD.

For the third vacuum at large $m$ the value $u=m^2/(- \det h)$
corresponds on the Coulomb branch to the so called charge vacuum, where
some fundamental fields become massless.  Moreover,  the correspondence
with \ntwo results can be demonstrated for the 
three vacua at any value of $m$. To
this end  we use the relation 
(\ref{ukappa}) and Eq.~(\ref{kappa}) to derive the
following equation for
$u$,
\begin{equation}
(-\det h)\,u^3-m^2\,u^2-\frac 9 8 \,(-\det
h)\,m\Lambda_1^3\,u+m^3\Lambda_1^3 +\frac{27}{2^8}\,(-\det
h)^2\Lambda_1^6=0\;.
\label{uone}
\end{equation}
The three roots of this equation are the vevs
 of $\Tr \,\Phi^2$ in the corresponding  vacua.

The equation (\ref{uone}) at  $\det h\!=\!-1$ coincides on the \ntwo 
side with the
condition of vanishing discriminant of the Seiberg-Witten curve,
\begin{equation}
y^2=x^3-u\,x^2+ \frac{1}{4}\Lambda_{1}^3 m\,x-\frac{1}{64}\Lambda_{1}^{6}\;.
\label{swu}
\end{equation}
Moreover, Eq.~(\ref{uone}) with $\det h\!\neq \!-1$ can be reduced to the case
$\det h\!=\! -1$ by  the rescaling
\begin{equation}
u= (-\det h)^{1/3}\, u'\,, \quad m= (-\det h)^{2/3}\, m'\,, 
\quad v= (-\det h)^{-1/3}\,v'\,.
\label{rescale}
\end{equation}
This is in agreement with the master parameter $\sigma$ which contains the
product $m^{-3/2}\det h$ and the nonanomalous U(1) symmetries
we  discussed
above. In other words, breaking of \ntwo by Yukawa couplings does not influence
consideration of the chiral condensates modulus the rescaling~(\ref{rescale}).

The consideration above shows that the only nonperturbative input needed
to determine  the chiral condensates
 is provided by the one-instanton ADS
superpotential. This  means that any reference
 to the \ntwo limit is not crucial at all,
i.e.  in regard to these condensates the exact Seiberg-Witten solution of \ntwo
is equivalent to the ADS superpotential.

The relations for the condensates we have
derived are not new, they were obtained in 
\cite{giveon} by the  `integrating in' procedure introduced in 
\cite{Intrilligator}.
Our approach which is based on `integrating out', plus the Konishi
relations, can  be 
viewed as an independent proof of the `integrating in' procedure.

What we see as an advantage of our approach it is that,
within a   certain range of
parameters, the superpotential (\ref{sup1}) gives a complete
 description of the low
energy physics.  Indeed, when the mass $m_V$ of the field $V$,
\begin{equation}
m_V= 2m\left(2-3\sigma \kappa\right)\;,
\end{equation}
is much less than the other masses, such as $m_\Phi=g^2\mu$ and $m_W=|g^2
v|^{1/2}$, we are in the weakly coupled Higgs phase and enjoy  full
theoretical
control. The Konishi relations help to determine the condensates of heavy
fields in  this
phase.  Holomorphy then allows for continuation of these
results for the condensates to strong coupling.

At strong coupling the superpotential 
(\ref{sup1}), like other versions of the Venezi\-ano-Yankielowicz
Lagrangians~\cite{vy}, 
does not describe the low energy physics. It can be viewed as a shorthand 
equation that gives the correct values of the condensates, an 
equivalent of `integrating in' procedure~\cite{giveon}.

One comment to add is about the photino condensate.
The gaugino condensate $\langle\Tr \lambda^2\rangle$ we found above
 can be viewed as a sum of the 
condensates for charged gauginos and the 
photino,
\begin{equation}
\left\langle\Tr \lambda^2\right\rangle=
\left\langle\lambda^+\lambda^-\right\rangle +
\frac 1 2\,
\left\langle\lambda^3\lambda^3\right\rangle
\end{equation}
In  gauge invariant form the photino condensate can be associated with\\ 
$
\left\langle(\Tr W \Phi)^2\right\rangle\,.
$.
It was shown in~\cite{HSZ}  that \ntwo is preserved in the effective  QED even 
when the breaking parameter $\mu$ is nonvanishing. An immediate consequence
of this observation is
 that the photino condensate vanishes, it is  {\em not} the lowest
component in the corresponding \ntwo supermultiplet. So,
 the gaugino condensate is solely due to the charged gluino.

\subsection{Argyres-Douglas points}

When the mass $m$  changes from large to small values we interpolate 
 between the two
quite different structures of vacua shown above. Let us consider this 
transition
when, for definiteness,
 \mbox{$\det h\!=\!-1$} and $m$ is real and positive
and changes 
  from large to small  values. 
 At large positive $m$ all the vacua are
situated at real values of  $u$, 
the dyon vacuum is at negative $u$, the monopole vacuum is at positive
$u$, and the
charge vacuum is also at positive, but much larger, values of $u$.  When $m$ 
diminishes then  at some point the  monopole and charge vacua collide on the
real axis of
$u$ and subsequently  go more off to  complex values
  producing the $Z_3$ picture at small $m$. 

The point in the parameter manifold where the two vacua coincide is the AD
 point~\cite{ad}. 
In the SU(2) theory these points were studied in \cite{apsw}.
Mutually non-local states, say charges and monopoles, becomes
massless at these points. On the Coulomb branch of the \ntwo theory
these points correspond to a non-trivial conformal field theory
\cite{apsw}. 

Here we study   the \none SUSY theory, where \ntwo is broken 
by the mass term for the adjoint matter as well as by of the
Yukawa coupling.   Collisions of two vacua 
still occur in this theory.  We  find the values of $m$
at which AD points appear  generalizing the
consideration in~\cite{apsw}.

Coalescence of two roots for $v$  means
that together with Eq.~(\ref{vaceq}) the derivative of its left-hand-side
should also vanish,
\begin{equation}
m-\frac{(-\det h)}{2} \, \frac{v}{\mu}- \frac{\Lambda_1^3}{4} \left(
\frac{\mu}{v}\right)^2=0, \qquad -(-\det h) + \Lambda_1^3 \left(
\frac{\mu}{v}\right)^3=0\;.
\label{vaceq1}
\end{equation}
This system is consistent only at three values of $m=m_{\rm AD}$,
\begin{equation}
m_{\rm AD}=\frac 3 4 \, \omega\, \Lambda_1\,(-\det  h)^{2/3},
\qquad \omega={\rm e}^{2\pi i n/3}\,,\qquad (n=0,\pm1)
\;,
\label{mad}
\end{equation}
related by $Z_3$ symmetry. The condensates at the AD vacuum are
\begin{eqnarray}
&&v_{\rm AD}=\omega \,\frac{\mu\,\Lambda_1}{(-\det
h)^{1/3}}\,,\qquad
u_{\rm AD}= \omega^{-1} \, \frac 3 4  \,  \Lambda_1^2\,  (-\det
h)^{1/3}\,,\nonumber\\[1mm]
&& s_{\rm AD}=\omega^{-1} \, \frac 1 4 \, \mu\,\Lambda_1^2 \,  (-\det h)^{1/3}\,.
\label{cad}
\end{eqnarray}

\section{Dyon condensates}\label{sec:mcd}

In this section we calculate various dyon condensates at the 
three vacua of the 
theory.
As  discussed above, holomorphy allows us to find these condensates
starting from  a consideration on the Coulomb branch in \ntwo near 
the singularities associated with a given massless dyon. Namely, 
we calculate the monopole
condensate
near the monopole point, the charge condensate near the charge point and the 
dyon
$(n_{m},n_{e})=(1,1)$ condensate near the point where this dyon is light. 
Although
we start with small values of the 
adjoint mass parameter $\mu$,  our results for
condensates are exact for any $\mu$ as well as for any value of $\det h$.

\subsection{Monopole condensate.}

Let us start with calculation of the monopole condensate near the
monopole point. Near this point the effective low energy
description of our theory can be given in terms of \ntwo dual QED~\cite{sw}. It
includes a light monopole hypermultiplet interacting with a 
vector (dual) photon
multiplet in the same way as electric charges interact with ordinary photons.
Following Seiberg and Witten~\cite{sw} we write down the effective 
superpotential
in the following form,
\begin{equation}
{\cal W}= \sqrt{2}\,\tilde{M}MA_{D}+\mu\, U,
\label{mqed}
\end{equation}
where $A_{D}$ is a neutral chiral  field (it is a part of the \ntwo
dual photon multiplet in the \ntwo theory) and $U\!\!=\!\Tr \Phi^2$
considered as a function of $A_{D}$. The
second  term 
breaks \ntwo supersymmetry down to \none.

Varying  this superpotential with respect to $A_{D}$, $M$
and $\tilde{M}$ we find that $A_{D}=0$, i.e.  the monopole mass vanishes, and
\begin{equation}
\langle\tilde{M}M \rangle=-\left.\frac{\mu}{\sqrt{2}}\frac{{\rm d}{u}\:}{{\rm
d}{a_{D}}}\right|_{a_{D}=0}\,.
\label{mc}
\end{equation}
The non-zero value of the monopole condensate $\langle\tilde{M}M
\rangle$ ensures  U(1) confinement for electric charges via the formation of
Abrikosov-Nielsen-Olesen  vortices. 

Let us work out
the r.h.s. of Eq.~(\ref{mc}) to determine the $\mu$
and $m$ dependence of the monopole condensate.
From the exact Seiberg-Witten solution \cite{sw}, we have
\begin{equation}
\frac{{\rm d}{a_{D}}}{{\rm d}u}=
\frac{\sqrt{2}}{8\pi }\oint_\gamma \frac{{\rm d}x}{y(x)}\,.
\label{con}
\end{equation}
Here for the Seiberg-Witten curve $y(x)$  given by Eq.~(\ref{swu}) we use 
the form
\begin{equation}
y^2=(x-e_{0})(x-e_{-})(x-e_{+})\,.
\end{equation}
The integration contour $\gamma$ in the $x$
plane  circles around  two branch points  $e_{+}$
and $e_{-}$ of $y(x)$. At the monopole vacuum, when  $u=u_{M}$, two
branch points 
$e_{+}$ and $e_{-}$ coincide, $e_{+}=e_{-}=e$ and the integral~(\ref{con}) is 
given
by the residue at $x=e$. 
\begin{equation}
\frac{{\rm d}{a_{D}}}{{\rm d}u}(u_{M}) =
\frac{i\,\sqrt{2}}{4\,\sqrt{e-e_0}}\,.
\end{equation}
The value of $e-e_0$ (equal at $u\!=\!u_M$ to $(1/2)\,{\rm d}^2 (y^2)/{\rm 
d}x^2$ )
is fixed by the equation ${\rm d}(y^2)/{\rm d}x=0$,
\begin{equation}
e-e_{0}=\sqrt{u_{M}^2-\frac 3 4 m\Lambda_{1}^{3}}\;.
\end{equation}
Substituting this into the expression for the monopole condensate
(\ref{mc})  we get finally
\begin{equation}
\langle \tilde M M\rangle =2i\mu\left(u_{M}^2-\frac 3 4
m\Lambda_{1}^{3}\right)^{1/4}.
\label{mm}
\end{equation}

To test the result let us consider first the limit of a 
large masses $m$ for the fundamental matter. 
 As  in Sec.~\ref{sec:sup} this limit can be viewed as a
RG flow to pure Yang-Mills theory with the identification
$
\Lambda_{0}^{4}=m\Lambda_{1}^3
$,
where $\Lambda_{0}$ is the scale of the \ntwo Yang-Mills theory.
In this theory we have $u_{M}=\Lambda_{0}^2$. Then Eq.~(\ref{mm})
gives
$
\langle \tilde M M\rangle =\sqrt{2}\,i\,\mu\,\Lambda_{0}\,,
$
which coincides with the Seiberg-Witten result~\cite{sw}. This ensures monopole
condensation and charge confinement in the monopole point at large $m$.

Notice, that in the derivation above \ntwo was not broken by the Yukawa 
coupling,
i.e. we assume $\det h\!=\!\!-1$. 
The result, however, can be easily generalized
to arbitrary 
$\det h$ by means  of U(1) symmetries considered above,
Eq.~(\ref{mm}) for the  monopole condensate remains
valid for arbitrary $\det h$.

\subsection{ Deconfinement in the Argyres-Douglas point}

Now let us address the question: what happens with the monopole
condensate when we reduce $m$ and approach the AD point?
The AD point corresponds to a particular
value of $m$ which ensures  coalescence of the monopole and charge
singularities in the $u$ plane. Near the monopole point we have
condensation of monopoles and confinement of charges while
near the charge point we have condensation of charges and
confinement of monopoles. As  shown by 't~Hooft 
these two phenomena cannot happen simultaneously~\cite{H}. The
question is:  what happens when monopole and charge points collide
in the $u$ plane?

The monopole condensate
 at the AD point is given by Eq.~(\ref{mm}). When $m$ and $u$ are
 substituted by 
 $m_{AD}$ and $u_{AD}$ from Eqs.~(\ref{mad}) and (\ref{cad}), we get
\begin{equation}
\langle \tilde M M\rangle_{AD}=0.
\end{equation}
We see that the monopole condensate goes to zero 
at the AD point. Our derivation makes it clear why it happens.
At the AD point all three roots of $y^2$
become degenerate, $e_+=e_-=e_0$, so the monopole condensate which is
proportional to $\sqrt{e-e_0}$ naturally vanishes.  

In the next subsection we
calculate the charge condensate in the charge point and show that it  also 
goes to
zero as $m$ approaches its AD value~(\ref{mad}). Thus, we interpret
the AD point as a deconfinement point for both monopoles
and charges.

\subsection{Charge and dyon condensates}

In this subsection we use the same method to calculate
values for the  charge and dyon condensates near the charge and dyon points
respectively. We first consider $m$ above its AD value (\ref{mad}) and
then continue our results to values of $m$ below $m_{AD}$. In
particular, in the limit $m=0$ we recover $Z_{3}$ symmetry.

Let us start with the charge condensate. At $\mu=0$, $\det h=\!-1$ and large
$m$  the
effective theory near the charge point 
\begin{equation}
a=-\sqrt{2}\,m
\label{chp}
\end{equation}
on the Coulomb branch
is \ntwo QED. Here $a$ is the neutral scalar, the partner of photon
in the \ntwo supermultiplet.
 Half of the degrees
of freedom in  color doublets  become massless whereas the other
half  acquire a large mass $2m$. The massless fields form one hypermultiplet
$\tilde Q_+, Q_+$ of charged particles in the effective electrodynamics.
 Once we add the mass term for the adjoint matter
 the effective superpotential near the charge point becomes
\begin{equation}
{\cal{W}}=\frac{1}{\sqrt{2}}\,\tilde Q_{+}Q_{+}A+m\,\tilde Q_{+}Q_{+}
 +\mu\, U 
\end{equation}
Minimizing this superpotential we get condition (\ref{chp})
as well as
\begin{equation}
\langle\tilde{Q}_{+}Q_{+}\rangle= -\left. \sqrt{2}\,\mu\,\frac{{\rm d} u}{{\rm d} a}\right|_{a=-\sqrt{2}\,m}\,.
\end{equation}
Now, following the same steps which led us from (\ref{mc}) to
 (\ref{mm}), we get
\begin{equation}
{\sqrt{-\det h}}\, \langle\tilde Q_{+}Q_{+}\rangle=2\,\mu\,
(u_{C}^2-\frac 3 4 
m\,\Lambda_{1}^{3})^{1/4}\,,
\label{chc}
\end{equation}
where we include a generalization to arbitrary $\det h$.
We choose to consider\\ ${\sqrt{-\det h}}\, \langle\tilde
Q_{+}Q_{+}\rangle$ because it has the U$_R$(1) charge equal to one,
similar to the  $\langle\tilde M M \rangle$ condensate considered above.
By $u_{C}$ we denote the position of the charge vacuum 
in the $u$ plane. 

Holomorphy allows us to extend the result (\ref{chc}) to arbitrary $m$ and
$\det h$. So we can use Eq.~(\ref{chc}) to find  the charge condensate at 
the AD
point. Using Eqs.~(\ref{mad}) and (\ref{cad}) we see that the 
charge condensate
vanishes at the AD point in the same manner the monopole condensate
does. As it was 
mentioned we interpret this as deconfinement for both charges and
monopoles.

To write results for the charge, monopole and dyon condensates together 
let us introduce the dyon field $D_{i}\,$, $i=1,2,3$, which
stands for the charge, monopole and $(1,1)$ dyon field,
\begin{equation}
D_{i}=\left\{(-\det h)^{1/4} Q_{+},\; M,\; D\right\}.
\end{equation}
The arguments of the previous subsection
which led us to the result (\ref{mm}) for monopole condensate
give for $\langle\tilde D_i \, D_i\rangle$ 
 \begin{equation}
\langle\tilde D_i \, D_i
\rangle=2\,i\,\zeta_{i}\,\mu
\left(u_{i}^2-\frac 3 4
\,m\,\Lambda_{1}^{3}\right)^{1/4},
\label{dyc}
\end{equation}
where $u_{i}$ is the position of the i-th point in the
$u$ plane and the  $\zeta_{i}$ are  phase factors. 

For the monopole condensate  at real values of $m$ larger than
the $m_{\rm AD}$  Eq.~(\ref{mm}) gives
$
\zeta_{M}=1,
$
while for the charge condensate from Eq.~(\ref{chc}) we have
$
\zeta_{C}=-i.
$
For the dyon the phase factor is 
$
\zeta_{D}= i\,.
$

At the particular AD point we have chosen the monopole and
charge  condensates vanish, while
the dyon condensate remains non-zero, see (\ref{dyc}). Below the AD point,
condensates are still given by  Eq.~(\ref{dyc}), but the
charge and monopole phase
factors  can change~\footnote{Note
that the quantum numbers of the ``charge'' and ``monopole'' are also
transformed, see \cite{BF}}. The dyon phase factor 
does not change when we move through the AD point  because the dyon
condensate does not vanish at this point.

In the limit $m=0$ we should recover the $Z_3$-symmetry for the
values of condensates. From Eq.~(\ref{dyc}) it is clear that the
absolute values of all three condensates are equal because
the values of the three roots $u_{i}$ are on the circle in the
$u$ plane, see (\ref{smc}). Imposing the requirement of
$Z_3$ symmetry at $m=0$ we can fix the unknown phase factors
$\zeta_{C}$  and $\zeta_{M}$ below the AD point using the value
$
\zeta_{D}= i
$
 for dyon.  This gives
$
\zeta_{C}= i\, ,\;\;
\zeta_{M}=- i\,.
$

\newpage 
\section{Conclusions}\label{sec:disc}

\begin{itemize}
\item 
In the \none theory the chiral condensates of matter and gaugino 
fields are fixed by the ADS superpotential as the only nonperturbative input. 
In the limit of small adjoint mass we find for condensates a complete
matching with  the \ntwo Seiberg-Witten solution. Although the bulk of our 
results
for matter and gaugino condensates overlaps with what is 
known in the literature we think that our
approach clarifies some aspects of duality in \none theories.
\item 
Using the Seiberg-Witten approach of the broken \ntwo we determine  the
monopole, charge, and dyon condensates in the \none theory.  
\item
The Argyres-Douglas points exist in the \none theories. 
When the mo\-no\-pole and charge vacua collide at the AD point both the monopole
and charge condensates vanish.  It results in the deconfinement of electric and
magnetic charges at the AD point.
\item
Vanishing of condensates signals existence of new nontrivial \none
super\-con\-for\-mal theories.
\end{itemize}


\newpage
\begin{thebibliography}{99}

\bibitem{GVY}
A.~Gorsky, A.~Vainshtein and A.~Yung,
Nucl.\ Phys.\  {\bf B584}, 197 (2000)
[hep-th/0004087].

\bibitem{ads}
I.~Affleck, M.~Dine and N.~Seiberg, Phys.\ Lett.\ {\bf B137}, 187
(1984).

\bibitem{SV}
V.A. Novikov, M.A. Shifman, A.I. Vainshtein, and V.I. Zakharov  
Nucl. Phys. {\bf B229}, 407 (1983) [Reprinted in {\em 
Supersymmetry}, Ed. S. Ferrara (North Holland/World Scientific, 
Amsterdam -- Singapore, 
1987), Vol. 1, page 606];
 M. Shifman and A. Vainshtein,  Nucl. Phys. {\bf B296}, 445 (1988);

\bibitem{IS}
K.~Intriligator and N.~Seiberg,
Nucl.\ Phys.\ Proc.\ Suppl.\  {\bf 45BC}, 1 (1996)
\mbox{[hep-th/9509066]}.

\bibitem{sw} 
N.~Seiberg and E.~Witten, 
 Nucl.~Phys. {\bf  B426}, 19 (1994); (E) {\bf B430}, 485 (1994) 
[hep-th/9407087];
 {\bf B431}, 484 (1994)  [hep-th/9408099].

\bibitem{kutasov}
D.~Kutasov, A.~Schwimmer and N.~Seiberg,
Nucl.\ Phys.\  {\bf B459}, 455 (1996)
[hep-th/9510222].

\bibitem{is}
K.~Intrilligator and N.~Seiberg,
Nucl.\ Phys.\  {\bf B431}, 551 (1994)
[hep-th/9408155]. 

\bibitem{giveon}
S.~Elitzur, A.~Forge, A.~Giveon, and E.~Rabinovici,
 Phys.\ Lett.\ {\bf B353}, 79 (1995) [hep-th/9504080];
Nucl.~Phys.\ {\bf B459},160 (1996) [hep-th/9509130]; 
\\
S. Elitzur, A. Forge, A. Giveon, K. Intrilligator, and  E. Rabinovici,
Phys.\ Lett.\ {\bf B379}, 121 (1996) [hep-th/9603051].

\bibitem{kitao}
S.~Terashima and S.~Yang,
 Phys.~Lett.\  {\bf B391}, 107 (1997) [hep-th/9607151];
T.~Kitao,
 Phys.~Lett.\ {\bf B402}, 290 (1997) \mbox{[hep-th/9611097]}.
T.~Kitao, S.~Terashima, and S.~Yang,
Phys.\ Lett.\ {\bf B399}, 75 (1997) 
\mbox{[hep-th/9701009]}.

\bibitem{kt}
K.~Konishi and H.~Terao,
Nucl.\ Phys.\  {\bf B511} (1998) 264
[hep-th/9707005].

\bibitem{ils}
K.~Intriligator, R.~G.~Leigh and N.~Seiberg,
Phys.\ Rev.\  {\bf D50}, 1092 (1994) \mbox{[hep-th/9403198]}.

\bibitem{Intrilligator}
K.~Intriligator,
Phys.\ Lett.\  {\bf B336}, 409 (1994)
[hep-th/9407106].

\bibitem{ad} 
P.~C.~Argyres and M.~R.~Douglas,
Nucl.\ Phys.\ {\bf B448}, 93 (1995) [hep-th/9505062].

\bibitem{apsw} 
P.~C.~Argyres, M.~R.~Plesser, N.~Seiberg and
E.~Witten, Nucl.\ Phys.\ {\bf B461}, 71 (1996) [hep-th/9511154].

\bibitem{hori}
T.~Eguchi, K.~Hori, K.~Ito and S.~Yang,
Nucl.\ Phys.\  {\bf B471}, 430 (1996)
\mbox{[hep-th/9603002]}.

\bibitem{DS}
M.~R.~Douglas and S.~H.~Shenker,
Nucl.\ Phys.\  {\bf B447}, 271 (1995)
[hep-th/9503163].

\bibitem{seibR}
N.~Seiberg,
Phys.\ Rev.\  {\bf D49}, 6857 (1994)
[hep-th/9402044].

\bibitem{NSVZ} 
V.~A.~Novikov, M.~A~ Shifman, A.~I.~Vainshtein
and V.~I.~Zakharov, Nucl.\ Phys.\ {\bf 260B}, 157 (1985).

\bibitem{ds}
G.~Dvali and M.~Shifman,
Phys.\ Lett.\  {\bf B396}, 64 (1997)
[hep-th/9612128].

\bibitem{vy}
G.~Veneziano and S.~Yankielowicz, Phys.\ Lett.\ {\bf B113}, 231 (1982); 
G.~Veneziano, S.~Yankielowicz and T.~Taylor, Nucl.\ Phys.\ {\bf B218}, 
493 (1983).

\bibitem{HSZ} 
A.~Hanany, M.~Strassler and A.~Zaffaroni,
 Nucl.\ Phys.\ {\bf B513}, 87 (1998) \mbox{[hep-th/9707244]}.

\bibitem{H} 
G.~'t Hooft,
Nucl.\ Phys.\ B {\bf 138}, 1 (1978);
Nucl.\ Phys.\ B {\bf 153}, 141 (1979).

\bibitem{BF}
A. Bilal and F. Ferrari, Nucl.\ Phys.\ {\bf B516}, 175 (1998) [hep-th/9706145].

\end{thebibliography}
\end{document}